# High-resolution cavity ring-down spectroscopy of the $\nu_1 + \nu_6$ combination band of methanol at 2.0 μm

**Hongming Yi[†] and Adam J. Fleisher[*]**

*Material Measurement Laboratory, National Institute of Standards and Technology, Gaithersburg, MD 20899*

[*]Correspondence to:   adam.fleisher@nist.gov

[†]Present address:   *Department of Civil and Environmental Engineering, Princeton University, Princeton, NJ 08544*

**Abstract:** Reported here are portions of the infrared absorption cross-section for methanol ($CH_3OH$) as measured by frequency-stabilized cavity ring-down spectroscopy (FS-CRDS) at wavelengths near $\lambda = 2.0$ µm. High-resolution spectra of two gravimetric mixtures of $CH_3OH$-in-air with nominal mole fractions of 202.2 µmol/mol and 45.89 µmol/mol, respectively, were recorded at pressures between 0.8 kPa and 102 kPa and at a temperature of 298 K. Covering the experimental wavenumber range of 4990 cm$^{-1}$ to 5010 cm$^{-1}$ in increments of 0.0067 cm$^{-1}$ and with an instrument linewidth of 30 kHz, we observed an evolution in the $CH_3OH$ spectrum from resolved absorption lines at a low pressure (0.833 kPa) to a pseudo-continuum of absorption at a near-atmospheric pressure (101.575 kPa). An analysis of resolvable features at the lowest recorded pressure yielded a minimum intramolecular vibrational energy redistribution (IVR) lifetime for the OH-stretch ($v_1$) plus OH-bend ($v_6$) combination of $\tau_{IVR} \geq 232$ ps — long compared to other methanol overtones and combinations. Consequently, we show that high-resolution FS-CRDS of this relatively weak $CH_3OH$ combination band provided an additional avenue by which to study the intramolecular dynamics of this simplest organic molecule with hindered internal rotation.

## I. INTRODUCTION

Methanol is central to our understanding of interstellar chemistry because its existence establishes a direct pathway to the formation of more complex organic molecules.[1] At low temperatures characteristic of dense interstellar clouds (<100 K), a potentially general tunneling mechanism is proposed to explain unexpectedly high reaction rates observed during the oxidation of organic molecules like methanol by hydroxyl radical.[2] Recently, pure rotational methanol transitions with coincidental cancellations of energies have also been proposed as a sensitive probe for variations in a fundamental constant.[3] Because of its importance in astrophysics, the calculation of accurate potential energy surfaces (and their derivative properties) for methanol using quantum-mechanical methods remains an active area of research.[4,5]

In the atmospheric sciences, spectral complexity makes difficult the creation of line-by-line models for radiative transfer in the presence of methanol. Quantum-mechanical spectral models for the low-frequency torsional bands near a wavelength of $\lambda = 10$ µm (Ref. 6) as well as for the CH-stretch region near $\lambda = 3.5$ µm (Ref. 7) are available, but are known to reproduce high-resolution infrared absorption cross-sections with only modest precision[8] compared to line-by-line models for other small molecules like carbon dioxide ($CO_2$).[9,10] Notwithstanding modeling challenges, methanol ($CH_3OH$) has been identified in the Earth's atmosphere,[11] as well as in cometary atmospheres[7,12,13] and in interstellar space.[14,15] Those formative findings paired with the pending deployment of satellite-based instruments with new observational capabilities in the infrared (e.g., Ref. 16) motivate further theoretical and experimental works towards the generation of accurate absorption cross-sections and line-by-line models for CH3OH with a variety of collisional partners.

More generally, methanol represents a case-study in small molecule internal dynamics because of its hindered internal rotation and low vibrational-mode-dependent torsional barrier.[17] Intramolecular vibrational energy redistribution (IVR) in the OH-stretch overtone spectrum was observed to occur on three distinct time scales,[18] as well as to depend upon the conformation of the coherently prepared upper state.[19] Such observations inspired new theoretical models for IVR using an adiabatic approximation,[20] as well as the development of double-resonance methods to explore state-specific IVR.[21]

Here we report frequency-stabilized cavity ring-down spectroscopy of the relatively weak $\nu_1 + \nu_6$ combination band near $\lambda = 2.0$ µm. Because methanol is known to exhibit mode-specific molecular dynamics and intramolecular vibrational energy redistribution (IVR),[22,23] we study the OH-stretch plus OH-bend ($\nu_1 + \nu_6$) combination band at high resolution, with a spectral sampling of 200.07 MHz (0.0067 cm$^{-1}$) and a precision of 30 kHz (1 × 10$^{-6}$ cm$^{-1}$). While the magnitude of the reported absorption cross-sections was only observed to be qualitative in nature due to variability in the outgassing of small molecules (e.g., H$_2$O, CO$_2$, CH$_3$OH, etc.) from the walls of our sample cell, we can report an approximate precision on the scaled cross-sections of ≈1 % — limited by achievable precision in the measured cavity ring-down time constants. In addition to reporting new high-resolution spectral data to bridge the gap between jet-cooled rotationally resolved spectra[22] and the available low-resolution spectroscopic reference data,[24] we report a simple analysis of observed line broadening which yields an estimated lower-bound for the IVR lifetime for the $\nu_1 + \nu_6$ combination band of ≥232 ps.

## II. EXPERIMENTAL

Frequency-stabilized cavity ring-down spectroscopy (FS-CRDS) of the $\nu_1 + \nu_6$ band of methanol (CH$_3$OH) was performed by temperature-tuning a series of distributed feedback diode lasers in a manner similar to that described in Yi et al.[25] The tuning range demonstrated here was increased to a wavenumber range of 4990 cm$^{-1}$ to 5010 cm$^{-1}$ by an additional laser, resulting in a twofold enhancement in spectrometer capabilities. For a given spectral acquisition, the infrared frequency of one of the continuous-wave lasers was tuned to be on resonance with a given mode of a high-finesse optical cavity of length $L$ = 75 cm containing a gas sample of CH$_3$OH-in-air from one of two gravimetric mixtures with CH$_3$OH mole fraction of either $\chi$ = 202.2 µmol/mol or $\chi$ = 45.89 µmol/mol. The spectral sampling (laser tuning step size) was defined by the cavity free spectral range of $\nu_{\text{fsr}} = \frac{c}{2L}$ = 200.07 MHz (where $c$ is the speed of light in a vacuum), a value measured by frequency-agile, rapid scanning spectroscopy[26] to a standard uncertainty of $u_{\text{fsr}}$ = 30 kHz (relative uncertainty of $u_{r,\text{fsr}}$ = 0.015 %). The cavity length (and therefore $\nu_{\text{fsr}}$) was passively stabilized by four invar rods which fortified the optical cavity length against drift and actively stabilized using a frequency-stabilized HeNe laser (ML-1, Micro-g LaCoste) with long-term frequency stability >2 × 10$^9$ per day and a piezoelectric actuator attached to one of the high-reflectivity CRDS mirrors.[27] The value of $u_{\text{fsr}}$ = 30 kHz (greater than the cavity line width of $\delta_{cav}$ = 4 kHz) was a measure of short-term jitter in the relative frequency of each cavity mode and thus a measure of the instrumental linewidth for high-resolution FS-CRDS.

By optically shuttering the laser injection at each discrete cavity mode using an acousto-optic modulator, the cavity time constant ($\tau$) for each cavity mode was measured in transmission using a digitizer with known nonideality[10] for nominally 100 consecutive cavity ring-down events at an acquisition rate of nominally 30 Hz. The relative standard uncertainties in $\tau$ at each frequency step were within the range from $u_{r,\tau}$ = 0.28 % to $u_{r,\tau}$ = 2.3 %, with a median value of $u_{r,\tau}$ = 0.78 % (approximately 11 000 unique frequency steps over four spectra). The absorption coefficient ($\alpha$) as a function of wavenumber ($\tilde{\nu}$) was calculated from the cavity time constants measured both with ($\tau$) and without ($\tau_0$) the CH$_3$OH-in-air sample:

$$\alpha(\tilde{\nu}) = \frac{1}{c\tau(\tilde{\nu})} - \frac{1}{c\tau_0(\tilde{\nu})}, \tag{1}$$

where $c$ is again the speed of light in a vacuum.

The sample cell and optical cavity were prepared to receive the CH$_3$OH-in-air sample by first introducing a purge of high-purity nitrogen (>99.99 %). Then, the sample cell was evacuated to a vacuum pressure (<1 Pa) prior to introducing the gas sample under study. Gas samples were maintained inside the cavity ring-down spectrometer at a temperature of $T$ = 298 K ± 1.3 K and measured over a series of pressures ranging from $p$ = 0.833 kPa to $p$ = 101.575 kPa (introducing a new sample each time the pressure was changed). In some instances, to further reduce the appearance of spectral interferences from additional molecules (e.g., water and carbon dioxide — see Section III.A), a gas sample of CH$_3$OH-in-air was left inside the optical cavity overnight, evacuated the next day, and immediately replaced with a CH$_3$OH-in-air sample for measurement.

Temperature was measured by a 100 Ω platinum resistance thermometer (Pt100/8A*G, Sensing Devices Incorporated) in good thermal contact with the outside of the sample cell, and temperature fluctuations from the environment were mitigated using an insulated box around the sample cell and optical cavity. Pressure was measured at the sample cell using a silicon resonant sensor digital manometer (MT210, Yokogawa) with full-scale range of 133 kPa. Standard uncertainties in $T$ and $p$ were traceable to the international system of units via secondary NIST standards and limited by systematic uncertainties to relative values of $u_{r,T} \leq 0.01$ % and $u_{r,p} \leq 0.004$ %, respectively.

The gas sample comprised a gravimetrically prepared mixture of methanol in air.[28] Briefly, methanol was introduced into an evacuated and pre-weighed 30 L gas cylinder via a syringe and septum. The syringe was weighed prior to and immediately following methanol injection into the evacuated cylinder, and the resulting mass difference constituted the delivered methanol amount of substance. The 30 L gas cylinder containing the injected methanol of known mass was then further diluted by air to a total pressure of approximately 12 MPa, and then weighed. For the gravimetrically prepared cylinders, the reported relative standard uncertainty in mole fraction was $u_{r,\chi} = 0.36$ %.[28]

## III. RESULTS AND DISCUSSIONS

Here we report the scaled FS-CRDS spectra of air-broadened methanol as progress towards an accurate infrared absorption cross-section within the atmospheric remote sensing window near λ = 2.0 μm.[11,29,30] Illustrated in Fig. 1 is a flow chart of the data analysis procedure which began with the experimental observable CRDS time constants and terminated with the scaled $CH_3OH$ infrared absorption cross-sections [$\sigma(\tilde{\nu})$]. Following the calculation of absorption coefficients using Eq. (1), spectral interferences associated with outgassing and wall exchange affects were removed from the $CH_3OH$ spectra using line-by-line reference data available in HITRAN2016.[31] Interference-corrected $\alpha(\tilde{v})$ were converted to a raw $\sigma(\tilde{v})$, and then scaled using two physical constraints. Firstly, the cross-section measured at atmospheric pressure was scaled to match the Pacific Northwest National Laboratory (PNNL) quantitative spectral database.[24] Then, each lower-pressure cross-section was constrained so that the integrated absorption coefficients were a linear function of the expected number density of absorbers as defined by the mole fraction of the gravimetric mixture and the measured total pressure. Finally, from scaled values of $\sigma(\tilde{v})$, scaled and interference-corrected values of $\alpha(\tilde{v})$ and $\tau(\tilde{v})$ could be calculated following the dashed arrows in Fig. 1.

Discussed in this section are details comprising the data analysis flow diagram, as well as an analysis of homogeneous line broadening affects and IVR. In brief, we report the treatment of spectral interferences from water ($H_2O$) and carbon dioxide ($CO_2$) (Section III.A), the scaling of the infrared absorption cross-sections using physical constraints (Sections III.B and III.C), the pressure dependence of the scaled infrared absorption cross-sections (Section III.D), and the estimated intramolecular dynamics determined from the observed linewidths in the high-resolution spectrum at low pressure (i.e., Doppler-broadened limit — see Section III.E). We note here that given our relatively narrow laser tuning range and the observed spectral complexity at $T$ = 298 K, the fitting of upper-state inertial parameters and the subsequent compilation of a line-by-line model for the air-broadened $v_1 + v_6$ band of methanol is beyond the scope of this work.

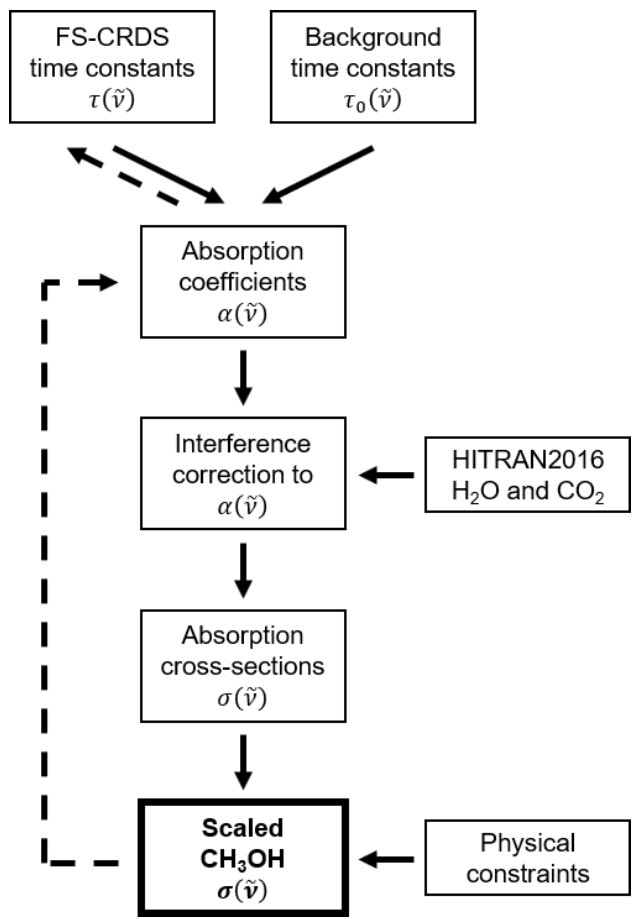

**Figure 1.** Data analysis flow diagram.

## A. Spectral interferences: $H_2O$ and $CO_2$

The experimental absorption spectra of $CH_3OH$-in-air were observed to include several interferences from $H_2O$ and $CO_2$ at wavenumbers $\tilde{v} < 4999.4$ cm$^{-1}$. The spectral interferences were only observed at $\tilde{v} < 4999.4$ cm$^{-1}$ — approximately half of the available spectral bandwidth — an observation attributed to an adjustment in the sample cell preparation and pretreatment which coincided with switching between laser diodes. As introduced in Section II, a high-purity $N_2$ purge and a low-pressure vacuum evacuation were performed prior to the introduction of the $CH_3OH$-in-air sample for all measurements. However, the presence of $H_2O$ and $CO_2$ absorption features at $\tilde{v} < 4999.4$ cm$^{-1}$ indicated that the two-step procedure was insufficient to eliminate outgassing and/or exchange with previously adsorbed molecules (particularly at pressures <10 kPa). Therefore, for subsequent measurements at $\tilde{v} \geq 4999.4$ cm$^{-1}$, we introduced a third sample cell preparation step: filling with a sacrificial CH$_3$OH-in-air sample for an overnight period (~14 h) prior to measurement. Each sacrificial sample was then evacuated and immediately replaced by a measurement sample, and the CH$_3$OH-in-air spectra at $\tilde{v} \geq 4999.4$ cm$^{-1}$ were recorded. The additional step apparently eliminated the major H$_2$O and CO$_2$ interferences, presumably by promoting exchange during the overnight period between the CH$_3$OH-in-air sample and previously adsorbed molecules.

We also suspect that outgassing from CH$_3$OH significantly influenced the number density of absorbers within our sample cell (again, especially at pressures <10 kPa). Anecdotally, we have observed outgassing from methanol adsorbed to the walls of our samples cell for weeks after cleaning the highly reflective CRDS mirrors with solvent. Therefore, scaling of the measured magnitude of each absorption cross-section using physical constraints (see Fig. 1 and Sections III.B and III.C) followed the removal of spectral interferences.

Plotted in Fig. 2a is the absorption spectrum of the 202.2 µmol/mol CH$_3$OH-in-air sample at a pressure of 5.490 kPa. The blue trace is the experimental spectrum derived from Eq. (1) and recorded both with ($\tilde{v} \geq 4999.4$ cm$^{-1}$) and without ($\tilde{v} < 4999.4$ cm$^{-1}$) the third sample cell preparation step (sacrificial sample). Plotted in red and black, respectively, are simulations of known H$_2$O and CO$_2$ absorption lines using reference data from the HITRAN2016 database.[31] The simulated H$_2$O and CO$_2$ absorption features match well with the strong lines in the experimental spectrum at $\tilde{v} < 4999.4$ cm$^{-1}$, but are notably absent in the experimental spectrum at $\tilde{v} \geq 4999.4$ cm$^{-1}$. For example, relatively strong H$_2$O lines (red trace) at 5004.5 cm$^{-1}$ and 5008 cm$^{-1}$ do not appear in the experimental data (blue trace).

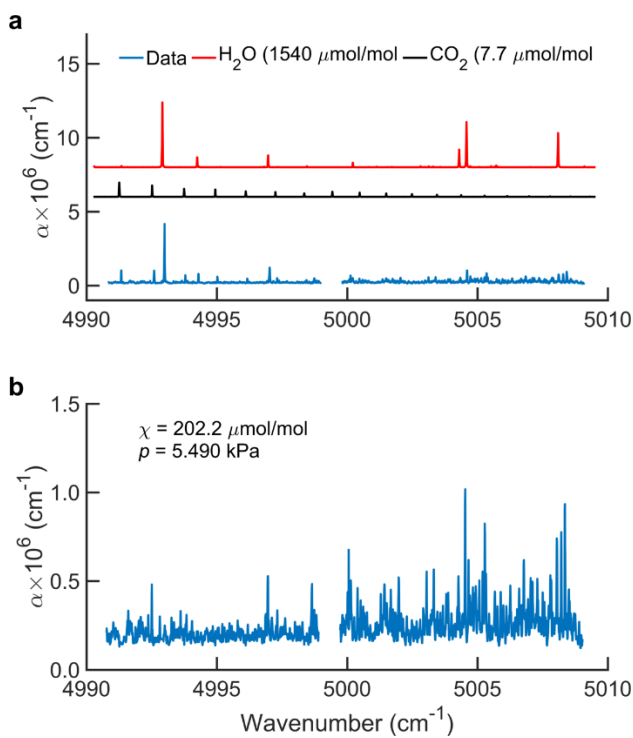

**Figure 2.** Interference analysis for the air-broadened spectrum of methanol (CH$_3$OH). The sample comprised 202.2 µmol/mol of CH$_3$OH-in-air at the pressure $p$ = 5.490 kPa. **a.** Simulated spectra of H$_2$O (red, offset by 8 × 10$^{-6}$ cm$^{-1}$) and CO$_2$ (black, offset by 6 × 10$^{-6}$ cm$^{-1}$) are plotted along with the raw experimental data (blue). Averaged mole fractions from the line-by-line analysis are listed in the legend. At $\tilde{v} \geq 4999.4$ cm$^{-1}$, the H$_2$O and CO$_2$ interferences were not observed in the raw data. **b.** The interference-corrected (and scaled) absorption spectrum of CH$_3$OH in air.

Spectral interferences observed at $\tilde{v} < 4999.4$ cm$^{-1}$ were removed from the experimental spectrum by varying the H$_2$O or CO$_2$ mole fraction for each interference feature. Line-by-line simulations were then subtracted from the experimental spectrum, resulting in an interference-corrected CH$_3$OH-in-air spectrum (Fig. 2b). Variability in the interference mole fractions was attributed to variability in outgassing as a function of time, temperature, and pressure, as well as to potential inaccuracies in the available reference data.

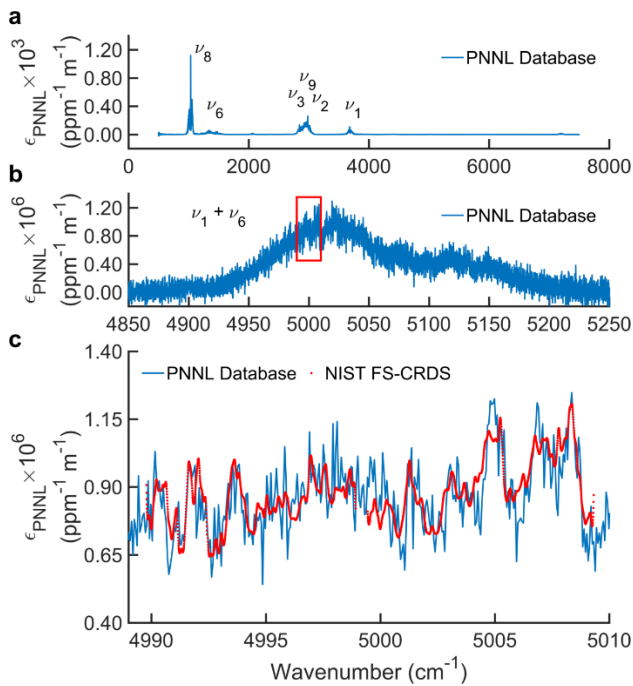

**Figure 3. a.** The Pacific Northwest National Laboratory (PNNL) quantitative spectral database for methanol. **b.** A 1000-fold vertical zoom in PNNL absorbance units ($\epsilon_{\text{PNNL}}$, ppm$^{-1}$ m$^{-1}$) at $\lambda = 2.0$ µm, revealing the weak $\nu_1 + \nu_6$ combination band. The National Institute of Standards and Technology (NIST) measurement window, limited by the tuning range of the available laser diodes, is enclosed by a red box. **c.** Comparison between PNNL Fourier transform spectroscopy (blue line) and the scaled NIST frequency-stabilized cavity ring-down spectroscopy (FS-CRDS, red dots) convolved with the PNNL instrument lineshape (ILS) function.

**B. Physical constraints I: PNNL spectral database**

The high-resolution FS-CRDS of methanol-in-air ($p = 101.575$ kPa, $\chi = 45.89$ µmol/mol) is compared to the Pacific Northwest National Laboratory (PNNL) quantitative spectral database (101.325 kPa ± 0.667 kPa)[24] in Fig 3. The entire PNNL database spectrum of methanol is plotted in Fig. 3a, and several of the fundamental vibrational bands are labeled therein. For convenience, a summary of the fundamental vibrational modes of methanol is available in Table 1.[21,32] In Fig. 3b, a 1000-fold vertical zoom near 5000 cm$^{-1}$ reveals the relatively weak $v_1 + v_6$ combination band previous assigned by jet-cooled rotationally resolved spectroscopy.[22] The red box illustrates the National Institute of Standards and Technology (NIST) FS-CRDS measurement window presented in this work, from 4990 cm$^{-1}$ to 5010 cm$^{-1}$.

TABLE 1: Fundamental vibrational frequencies of methanol.[a]

| vibrational state | label | experimental gas-phase wavenumber (cm$^{-1}$)[b] | description |
|---|---|---|---|
| $v_1$ | $v$ (OH) | 3682 | OH-stretch |
| $v_2$ | $v$ (CH)$_a$ | 3004 | asymmetric CH-stretch |
| $v_3$ | $v$ (CH)$_s$ | 2845 | symmetric CH-stretch |
| $v_4$ | $\delta$ (CH)$_a$ | 1484 | asymmetric CH-bend |
| $v_5$ | $\delta$ (CH)$_s$ | 1453 | symmetric CH-bend |
| $v_6$ | $\delta$ (OH) | 1335 | OH-bend |
| $v_7$ | $\rho_{\parallel}$ (CH$_3$) | 1070 | parallel CH$_3$-rock |
| $v_8$ | $v$ (CO) | 1034 | CO-stretch |
| $v_9$ | $v$ (CH)$_a$ | 2961 | asymmetric CH-stretch |
| $v_{10}$ | $\delta$ (CH)$_a$ | 1474 | asymmetric CH-bend |
| $v_{11}$ | $\rho_{\perp}$ (CH$_3$) | 1156 | perpendicular CH$_3$-rock |
| $v_{12}$ | -- | 200 | torsion |

[a]Adapted from Table 1 of Twagirayezu et al.[21]
[b]Similar values[32] are also listed in the NIST Chemistry WebBook.

Due to severe limitations in our ability to accurately know the true number density of CH$_3$OH in our sample cell, we chose to scale our qualitative high-resolution FS-CRDS to match the PNNL database. Firstly, we converted our interference-corrected $\alpha(\tilde{v})$ in units of cm$^{-1}$ to PNNL units of base-10 absorbance per part-per-million of the absorber per meter of pathlength (ppm⁻¹ m⁻¹). Following the formalism in Harrison and Bernath[33] we converted from the FS-CRDS spectral transmittance [$\mathcal{T}(\tilde{v})$] to PNNL decadal absorbance units ($\epsilon_{\text{PNNL}}$ in ppm⁻¹ m⁻¹) using Eqs. (2)-(3).

$$\epsilon_{\text{PNNL}} = -\zeta \frac{T}{296} \frac{0.101325}{\chi p L_{\text{eff}}} \log_{10} \mathcal{T}$$

$$= \zeta \frac{0.101325}{296 \times 10^4 k_B \ln(10)} \sigma(\tilde{v}) \qquad (2)$$

Note that the numerical factor (excluding our scaling parameter, $\zeta$) to convert from the absorption cross-section [$\sigma(\tilde{v})$] in units of cm² molecule⁻¹ to $\varepsilon_{\text{PNNL}}$ in absorbance units of ppm⁻¹ m⁻¹ is ~9.286 97 × 10⁻¹⁶. The scaling parameter ($\zeta$) was fitted to match our qualitative cross-section to the PNNL database, where the absorption cross-section as defined in Kochanov et al.[34] is

$$\sigma(\tilde{v}) = \frac{-\ln\{\mathcal{T}(\tilde{v})\}}{\chi \rho L_{\text{eff}}} = \frac{\alpha(\tilde{v})}{\chi \rho}. \qquad (3)$$

In Eq. (3), $\mathcal{T}(\tilde{v}) = \exp\{-\alpha(\tilde{v}) L_{\text{eff}}\}$ is again the spectral transmittance, $\alpha(\tilde{v})$ is the absorption coefficient from Eq. (1) as a function of wavenumber, $L_{\text{eff}} = FL/\pi$ is the effective cavity-enhanced path length for CRDS,[35] $F$ is the finesse of the optical cavity, $L$ is the length of the optical cavity, $\rho = \frac{p}{k_B T}$ is the number density of the gas sample, $k_B$ is the Boltzmann constant, and $\chi$ is again the mole fraction of absorbers. Finally, the scaled absorption cross-section at atmospheric pressure is defined here as $\zeta \sigma(\tilde{v})$.

In Fig. 3c, the NIST FS-CRDS measurement (red dots) converted to $\epsilon_{\text{PNNL}}$ is overlapped with the PNNL database (blue line). The NIST spectrum shown in Fig. 3c was convolved with the PNNL instrument lineshape (ILS) function reported in the methanol spectrum metadata to be a normalized sinc function [$2\Delta\text{sinc}(2\pi\tilde{v}\Delta)$][36] with full-width at half-maximum (FWHM) resolution $\Delta\tilde{v}_{\text{ILS}} = 0.73/\Delta = 0.112$ cm⁻¹ limited by the maximum zero path difference of the Fourier transform infrared spectrometer.[24] [Note that convolution with the PNNL ILS function had a minor effect on the appearance of the NIST spectrum compared to that reported in Section III.D, most likely due to collisional broadening estimated to be of the same order-of-magnitude as $\Delta\tilde{\nu}_{ILS}$ ($\Delta\tilde{\nu}_{col} = 2\gamma_{air}p \approx 0.2$ cm$^{-1}$, where $\gamma_{air}$ is the half-width at half-maximum air-broadening coefficient[31] — see Section III.E).] The high-resolution FS-CRDS reported, while scaled in magnitude to match the PNNL database, does show significantly improved measurement precision as indicated by the reduced spectral noise.

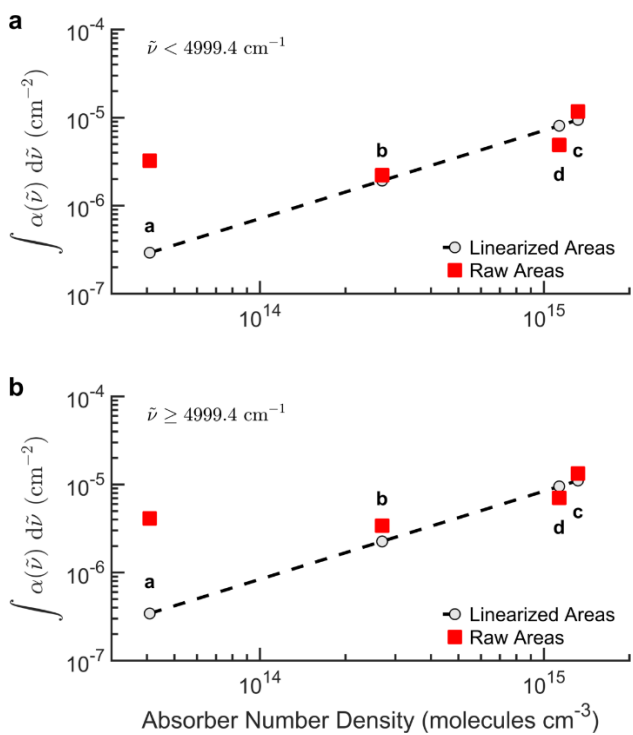

**Figure 4.** Linear relationship between the integrated absorption coefficient [$\int \alpha(\tilde{\nu})d\tilde{\nu}$] and the expected number density of CH$_3$OH absorbers ($\chi\rho$) for **a.** $\tilde{\nu} < 4999.4$ cm$^{-1}$ and **b.** $\tilde{\nu} \geq 4999.4$ cm$^{-1}$. Lowercase letters in both panels correspond to the scaled absorption cross-sections labeled in Fig. 5 of Section III.D.

## C. Physical constraints II: spectral area vs. number density of absorbers

With the absorption cross-section measured at near-atmospheric pressure ($p$ = 101.575 kPa) scaled to match the PNNL database, we further constrained the pressure-dependent absorption cross-sections by forcing the area under each absorption spectra to scale linearly with the expected number density of absorbers ($\chi\rho$). From Eqs. (2)-(3), the integrated absorption cross-section [$S_\sigma \equiv \int \sigma(\tilde{v})d\tilde{v}$] is equal to the integrated absorption coefficient scaled by the inverse of the number density of absorbers. Therefore, we write

$$\chi\rho S_\sigma = \zeta \int \alpha(\tilde{v})d\tilde{v} \qquad (4)$$

were the integration would ideally be performed over the entire absorption band and $\zeta$ are again the scaling parameters. The linear relationship for the partial cross-section is shown in Fig. 4, were each scaled data point (gray circles) is labeled with a lowercase letter to match its respective scaled absorption cross-section in Fig. 5 of Section III.D. Also shown in Fig. 4 are the integrated absorption coefficients prior to scaling (red squares). Clearly, the number density of methanol in our gas sample was not accurately reproduced by $\chi\rho$, particularly at the lowest pressure of $p$ = 0.833 kPa. We attribute this large discrepancy at low pressure to outgassing of methanol from prior mirror cleaning treatments. However, further investigation is required to confirm this uncharacterized interference. A summary of the scaling factors applied to each absorption cross-section is presented in Table 2.

## D. Scaled infrared absorption cross-sections of the $v_1 + v_6$ band

Infrared absorption cross-sections measured at a variety of temperatures and pressures are increasingly compiled into spectroscopic databases, especially for large molecules with complicated spectra. Their usefulness in radiative transfer codes (e.g., Kochanov et al.[34]) has recently motivated an expansion of HITRAN2016 to included approximately 300 cross-section files.[31] However, no cross-sections files for methanol at $\tilde{v} \geq 3250$ cm$^{-1}$ are currently available in HITRAN2016.

TABLE 2: Methanol absorption cross-section scaling parameters ($\zeta$).

| $\chi^a$ (μmol/mol) | $p$ (kPa) | $T$ (K) | $\zeta$ ($\tilde{v} < 4999.4$ cm$^{-1}$) | $\zeta$ ($\tilde{v} \geq 4999.4$ cm$^{-1}$) |
|---|---|---|---|---|
| 202.2 | 0.833 | 298.4 | 0.09 | 0.08 |
| 202.2 | 5.490 | 298.5 | 0.86 | 0.66 |
| 202.2 | 26.859 | 298.2 | 0.80 | 0.83 |
| 45.89 | 101.57 | 298.1 | 1.65 | 1.35 |

$^a$Mole fraction of the gravimetrically prepared gas sample.

The scaled high-resolution FS-CRDS cross-sections of methanol are plotted in Fig. 5. In each panel, a small gap in the combined laser coverage is visible at 4999.4 cm$^{-1}$. In the top two panels recorded at lower pressure, resolved methanol lines are visible. At higher pressure (bottom two panels), the distinct features become blended as pressure-broadening presumably merges lines within the congested combination band.

### E. Estimated IVR dynamics

Following the hypothetical dynamics treatment discussed in Rueda et al.,[22] we calculated a lower-bound for the IVR lifetime ($\tau_{\text{IVR}}$) of the $v_1 + v_6$ combination band of methanol while making the following assumptions: 1) all resolved features with peak cross-section $\sigma_i \geq 5 \times 10^{-22}$ cm$^2$ molecule$^{-1}$ were isolated spectral lines (i.e., unique optical transitions between two eigenstates), 2) all line-broadening in excess of the calculated

Doppler-broadening FWHM ($\Delta\tilde{\nu}_D = 0.011$ cm$^{-1}$) and the estimated air-broadened collisional FWHM ($\Delta\tilde{\nu}_{col} = 0.0016$ cm$^{-1}$ at $p = 0.833$ kPa)[31] was attributable to homogeneous broadening from IVR, and 3) the cross-sections were properly normalized, interference-correction and scaled using Eqs. (1)-(4) and the physical constraints discussed in Sections III.B and III.C. Those assumptions, which ignored spectrally overlapping transitions, yielded an upper-bound estimate for homogeneous broadening from IVR, and therefore a lower-bound value of $\tau_{IVR}$.

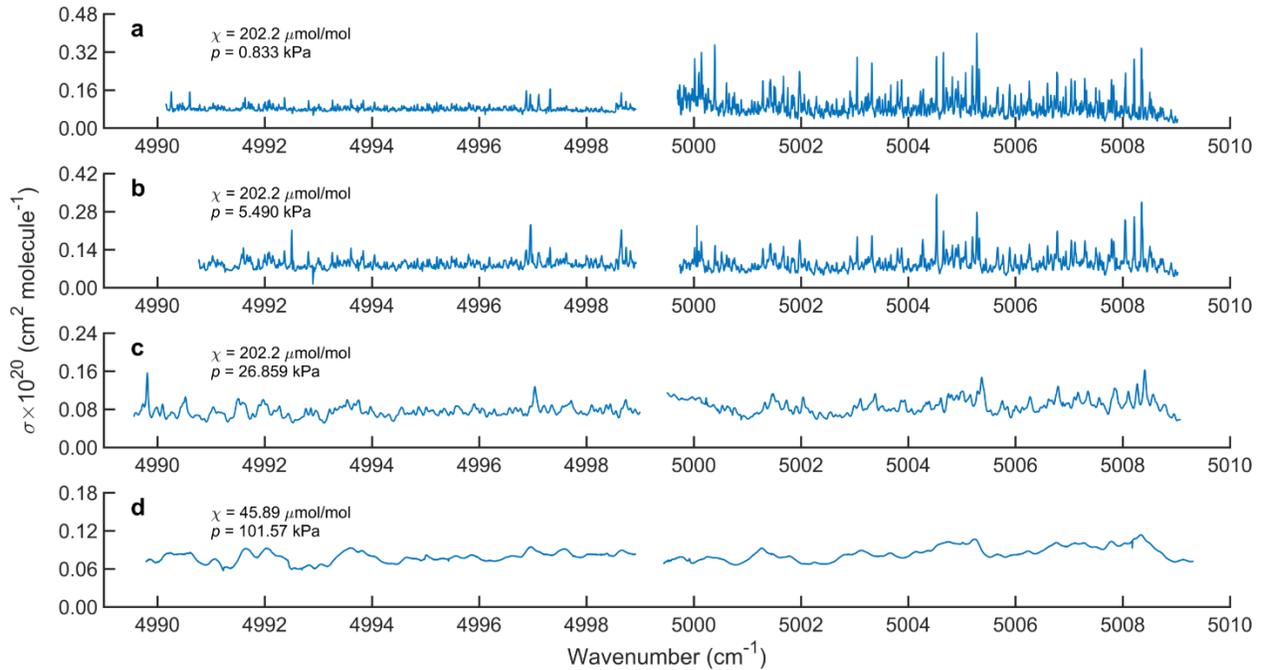

**Figure 5.** Infrared absorption cross-sections for the $\nu_1 + \nu_6$ band of methanol at a temperature of $T = 298$ K. The sample comprised 202.2 µmol/mol of CH$_3$OH-in-air at $p$ equal to **a.** 0.833 kPa, **b.** 5.490 kPa, **c.** 26.859 kPa. Shown in **d.** is a sample of 45.89 µmol/mol CH$_3$OH-in-air at $p = 101.575$ kPa.

The Doppler-broadened FWHM was calculated using the well-known equation $\Delta\tilde{\nu}_D = \sqrt{\frac{8N_A k_B T \ln 2}{mc^2}}\,\tilde{\nu}_0$, where $N_A$ is the Avogadro constant, $k_B$ is the Boltzmann constant, $m$ is the molecular molar mass in units of kg/mol, and $\tilde{\nu}_0$ is the transition wavenumber ($\tilde{\nu}_0 \approx 5000$ cm$^{-1}$). The air-broadened collisional FWHM was estimated using the air-broadening half-width coefficient of $\gamma_{\text{air}} = 0.1$ cm$^{-1}$ atm$^{-1}$ available in HITRAN for several fundamental bands of methanol near 1030 cm$^{-1}$ beginning in 2004.[38] Applying an automated peak-finding and analysis algorithm in MATLAB software to the low-pressure scaled cross-section shown in Fig. 5a ($p = 0.833$ kPa) using a minimum peak prominence of $5 \times 10^{-22}$ cm$^2$ molecule$^{-1}$, we identified 134 resolved features with FWHM $\Delta\tilde{\nu} \leq 20\Delta\tilde{\nu}_{\text{D}}$. The machine-picked values of $\Delta\tilde{\nu}$ are plotted in Fig. 6a, along with a dashed line identifying the $\Delta\tilde{\nu}_{\text{D}}$ low-pressure limit.

The distribution of $\tau_{\text{IVR}} = 1/(2\pi c \Delta\tilde{\nu}_{\text{L}})$ calculated from the excess homogeneous broadening $\Delta\tilde{\nu}_{\text{L}}$ is shown in Fig. 6b. The excess broadening ($\Delta\tilde{\nu}_{\text{L}}$), equal to the total homogeneous broadening minus the collisional broadening ($\Delta\tilde{\nu}_{\text{L}} = \Delta\tilde{\nu}_{\text{H}} - \Delta\tilde{\nu}_{\text{col}}$), was calculated using an empirical approximation (1 % accuracy) for the FWHM of a Voigt function shown in Eq. (5).[37]

$$\Delta\tilde{\nu} = \frac{\Delta\tilde{\nu}_{\text{H}}}{2} + \sqrt{\frac{\Delta\tilde{\nu}_{\text{H}}^2}{4} + \Delta\tilde{\nu}_{\text{D}}^2} \qquad (5)$$

Solving Eq. (5) for $\Delta\tilde{\nu}_{\text{H}}$ and using the substitution $\Delta\tilde{\nu}_{\text{H}} = \Delta\tilde{\nu}_{\text{L}} + \Delta\tilde{\nu}_{\text{col}}$ resulted in an approximate expression for the excess homogenous broadening which we attribute to IVR: $\Delta\tilde{\nu}_{\text{L}} = \frac{\Delta\tilde{\nu}^2 - \Delta\tilde{\nu}_{\text{D}}^2}{\Delta\tilde{\nu}} - \Delta\tilde{\nu}_{\text{col}}$.

In our simple analysis, we take the mean value of a lognormal distribution function fitted to the histogram data in Fig. 6b (red dashed line) to be a lower-bound for the IVR lifetime, $\tau_{\text{IVR}} \geq 232$ ps. Variations in the chosen peak-picking conditions as well as the statistical uncertainty in the distribution mean value yielded a standard uncertainty in the lower-bound value of $\sigma_{\tau_{IVR}} \approx 24$ ps. Therefore, our final estimated IVR lifetime of the $\nu_1 + \nu_6$ band of methanol is $\tau_{IVR} \geq 232$ ps ± 24 ps.

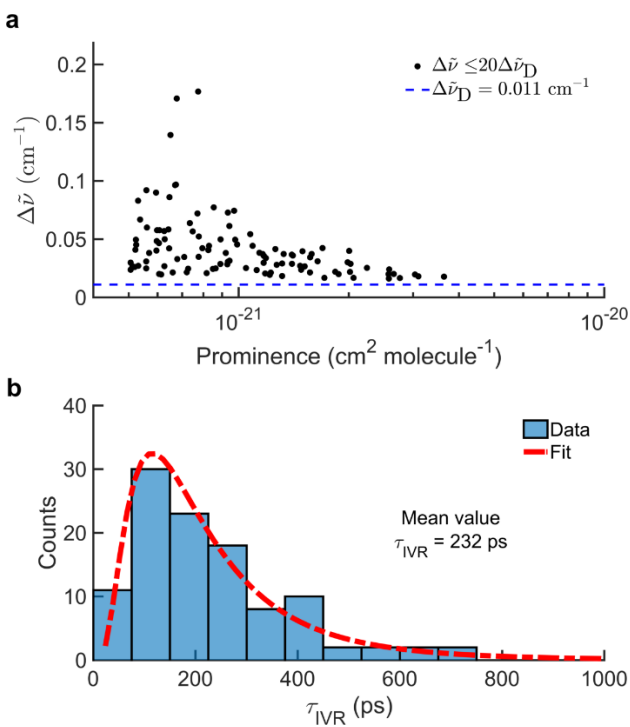

**Figure 6.** Analysis of resolvable line broadening in the low-pressure ($p = 0.833$ kPa) spectrum of CH₃OH-in-air ($\chi = 202.2$ µmol/mol) near 2.0 µm ($\nu_1 + \nu_6$ combination band). **a.** Scatter-plot of resolved feature full-widths at half-maximum (FWHM, $\Delta\tilde{\nu}$) versus peak prominence (black dots), along with the calculated Doppler-broadened FWHM of $\Delta\tilde{\nu}_D = 0.011$ cm⁻¹ (blue dashed line). **b.** Histogram of $\tau_{IVR} = 1/(2\pi c \Delta\tilde{\nu}_L)$, where $\Delta\tilde{\nu}_L = \frac{(\Delta\tilde{\nu}^2 - \Delta\tilde{\nu}_D^2)}{\Delta\tilde{\nu}} - \Delta\tilde{\nu}_{col}$ and $c$ is the speed of light. Also plotted is a fitted lognormal distribution function (red dashed line) with mean value of $\tau_{IVR} = 232$ ps.

The IVR dynamics analysis assumed that the contributions to pressure broadening from collisional with partners other than the bath gas of air (e.g., H₂O, CO₂, CH₃OH, etc.) were negligible, and therefore that the scaling using physical constraints was simply a correction for the unknown number density of absorbers. In other words, we assumed that the scaling using physical constraints did not affect the fidelity of the reported frequency axis, and that the pressure-broadening contribution to the homogeneous line width was independent of $\chi\rho$. At true mole fractions of CH₃OH-in-air of $\chi/\zeta \leq 0.25$ % (where $\zeta$ are the scaling parameters in Table 2), the contribution of self-broadening to the homogeneous linewidth analysis was no larger than the same order-of-magnitude as the estimated relative statistical uncertainty in $\tau_{\text{IVR}}$ (~0.1 %).

Finally, we reiterate that pressure broadening coefficients ($\gamma_{\text{air}}$, $\gamma_{\text{self}}$) for methanol are generally unknown, with the aforementioned values of $\gamma_{\text{air}} = 0.1$ cm$^{-1}$ atm$^{-1}$ appearing in HITRAN for only some strong infrared bands beginning in 2004.[38] At $p = 0.833$ kPa (Fig. 1a), the assumed pressure broadening coefficient yielded the collisional FWHM of $\Delta\tilde{\nu}_{\text{col}} = 0.0016$ cm$^{-1}$, and therefore an estimated collisional lifetime of $\tau_{\text{col}} \approx 3$ ns.

## IV. CONCLUSIONS

Here we report frequency-stabilized cavity ring-down spectroscopy (FS-CRDS) of the $\nu_1 + \nu_6$ band of methanol (CH₃OH) at a wavelength of $\lambda = 2.0$ μm. The high-resolution spectroscopy yielded portions of the infrared absorption cross-section which were then scaled using reasonable physical constraints. This work towards accurate line-by-lie reference data for methanol — the simplest organic alcohol known to exist in the Earth's atmosphere as well as in cometary atmospherics and interstellar space — reinforced the difficulties associated with performing quantitative spectroscopy on polar molecules at low mole fraction. Experimentally, the wavelength coverage of the reported FS-CRDS instrument was improved twofold by the addition of a new laser diode, thus enabling preliminary explorations of broadly absorbing species like methanol which exhibit complex spectra due to hindered internal rotation. We are currently working to replace the diode lasers entirely with a broadly tunable external cavity diode laser (4820 cm$^{-1}$ to 5060 cm$^{-1}$), consequently enabling an even broader survey of the absorption spectroscopy of CH$_3$OH near $\lambda$ = 2.0 µm.

The estimated intramolecular vibrational energy redistribution (IVR) lifetime resulting from our analysis of the room-temperature ($T$ = 298 K) methanol spectrum is consistent with the qualitative differences in the jet-cooled spectra observed by Rueda et al.,[22] where combinations of the OH-stretch ($\nu_1$) were reported to exhibited narrow features (e.g., 0.1 cm$^{-1}$ for the $2\nu_1 + \nu_8$ band). The narrowest features observed by Rueda et al. were approximately double their nominal dye laser linewidth of 0.05 cm$^{-1}$,[22] suggesting that the estimated IVR lifetime of ≈50 ps for bands involving the OH-stretch ($\nu_1$) was influenced by the instrument line shape function. With an instrument linewidth of $1 \times 10^{-6}$ cm$^{-1}$ (30 kHz), we report a lower-bound IVR lifetime of $\tau_{IVR} \geq 232$ ps as further confirmation that IVR in methanol is vibrational-mode-dependent, and indeed slow for the OH-stretch plus OH-bend ($\nu_1 + \nu_6$) combination band. Future studies of the pressure-dependent homogeneous broadening for the OH-stretch combination bands of methanol should further reveal compelling evidence of mode-specific IVR dynamics.

**SUPPLEMENTARY MATERIAL**

See supplementary material for data files containing the air-broadened, interference-corrected and scaled infrared absorption cross-section for methanol at $\lambda$ = 2.0 µm plotted in Fig. 5.

**ACKNOWLEDGMENT**

We are grateful to Thomas R. Rizzo (École Polytechnique Fédérale de Lausanne, EPFL, Switzerland) and David S. Perry (University of Akron) for sharing the jet-cooled rotationally resolved spectrum of methanol reported in Fig. 2 of Rueda et al.[22] which aided in our initial band assignment. We are also grateful to Timothy J. Johnson (Pacific Northwest National Laboratory, PNNL) for sharing the methanol spectrum and associated metadata from the PNNL database, and to George C. Rhoderick (National Institute of Standards and Technology, NIST) for discussing the gas metrology of the $CH_3OH$-in-air samples. We acknowledge William S. McGivern (NIST) and Fabrizio R. Giorgetta (NIST) for commenting on the manuscript. This research was supported by NIST.

Certain commercial equipment and software is identified in this paper in order to specify the experimental procedure adequately. Such identification is not intended to imply recommendation or endorsement by NIST, nor is it intended to imply that the equipment or software identified is necessarily the best available for the purpose.

**REFERENCES**

1. D. C. B. Whittet, A. M. Cook, E. Herbst, J. E. Chiar, and S. S. Shenoy, "Observational constraints on methanol production in interstellar and preplanetary ices," *Astrophys. J.* **742,** 28 (2011).

2. R. J. Shannon, M. A. Blitz, A. Goddard, and D. E. Heard, "Accelerated chemistry in the reaction between the hydroxyl radical and methanol at interstellar temperatures facilitated by tunneling," *Nat. Chem.* **5,** 745-749 (2013).

3. P. Jansen, L.-H. Xu, I. Kleiner, W. Ubachs, and H. L. Bethlem, "Methanol as a sensitive probe for spatial and temporal variations of the proton-to-electron mass ratio," *Phys. Rev. Lett.* **106,** 100801 (2011).

4. C. Qu, and J. M. Bowman, "Full-dimensional, *ab initio* potential energy surface for $CH_3OH \rightarrow CH_3 + OH$," *Mol. Phys.* **111,** 1964-1971 (2013).

5. A. Nandi, C. Qu, and J. M. Bowman, "Diffusion Monte Carlo calculations of zero-point energies of methanol and deuterated methanol," *J. Comput. Chem.* **40,** 328-332 (2019).

6. L.-H. Xu, R. M. Lees, P. Wang, L. R. Brown, I. Kleiner, and J. W. C. Johns "New assignments, line intensities, and HITRAN database for $CH_3OH$ at 10 µm," *J. Mol. Spectrosc.* **228,** 453-470 (2004).

7. G. L. Villanueva, M. A. DiSanti, M. J. Mumma, and L.-H. Xu, "A quantum band model of the $\nu_3$ fundamental of methanol ($CH_3OH$) and its application to fluorescence spectra of comets," *Astrophys. J.* **747,** 37 (2012).

8. J. J. Harrison, N. D. C. Allen, and P. F. Bernath, "Infrared absorption cross sections for methanol," *J. Quant. Spectrosc. Radiat. Transfer* **113,** 2189-2196 (2012).

9. F. Oyafuso, V. H. Payne, B. J. Drouin, V. M. Devi, D. C. Benner, K. Sung, S. Yu, I. E. Gordon, R. Kochanov, Y. Tan, D. Crisp, E. J. Mlawer, and A. Guillaume, "High accuracy absorption coefficients for the Orbiting Carbon Observatory-2 (OCO-2) mission: Validation of updated carbon dioxide cross-sections using atmospheric spectra," *J. Quant. Spectrosc. Radiat. Transfer* **203,** 213-223 (2017).

10. A. J. Fleisher, E. M. Adkins, Z. D. Reed, H. Yi, D. A. Long, H. M. Fleurbaey, and J. T. Hodges, "Twenty-five-fold reduction in measurement uncertainty for a molecular line intensity," *Phys. Rev. Lett.* **123,** 043001 (2019).

11. G. Dufour, C. D. Boone, C. P. Rinsland, and P. F. Bernath, "First space-borne measurements of methanol inside aged southern tropical to mid-latitude biomass burning plumes using the ACE-FTS instrument," *Atmos. Chem. Phys.* **6,** 3463-3470 (2006).

12. D. Bockelée-Morvan, P. Colom, J. Crovisier, D. Despois, and G. Paubert, "Microwave detection of hydrogen sulphide and methanol in comet Austin (1989c1)," *Nature* **350,** 318-320 (1991).

13. S. Hoban, M. J. Mumma, D. C. Reuter, M. DiSanti, R. R. Joyce, and A. Storrs, "A tentative identification of methanol as the progenitor of the 3.52-µm emission feature in several comets," *Icarus* **93,** 122-134 (1991).

14. J. A. Ball, C. A. Gottlieb, A. E. Lilley, and H. E. Radford, "Detection of methyl alcohol in Sagittarius," *Astrophys. J.* **162,** L203 (1970).

15. L. E. B. Johansson, C. Andersson, and J. Ellder, "Spectral scan of Orion A and IRC+10216 from 72 to 91 GHz," *Astron. Astrophys.* **130,** 227-256 (1984).

16. S. N. Milam, J. A. Stansberry, G. Sonneborn, and C. Thomas, "The *James Webb Space Telescope*'s plan for operations and instrument capabilities for observations in the Solar System," *Publ. Astron. Soc. Pac.* **128,** 018001 (2016).

17. R. Pearman and M. Gruebele, "Approximate factorization of molecular potential surfaces II. Internal rotors," *Z. Phys. Chem.* **214,** 1439 (2000).

18. O. V. Boyarkin, T. R. Rizzo, and D. S. Perry, "Intramolecular energy transfer in highly vibrationally excited methanol. II. Multiple time scale of energy transfer," *J. Chem. Phys.* **110,** 11346 (1999).

19. P. Maksyutenko, O. V. Boyarkin, and T. R. Rizzo, "Conformational dependence of intramolecular vibrational redistribution in methanol," *J. Chem. Phys.* **126,** 044311 (2007).

20. T. N. Clasp and D. S. Perry, "Torsion-vibration coupling in methanol: The adiabatic approximation and intramolecular vibrational redistribution scaling," *J. Chem. Phys.* **125,** 104313 (2006).

21. S. Twagirayezu, T. N. Clasp, D. S. Perry, J. L. Neill, M. T. Muckle, and B. H. Pate, "Vibrational coupling pathways in methanol as revealed by coherence-converted population transfer Fourier transform microwave infrared double-resonance spectroscopy," *J. Phys. Chem. A* **114,** 6818-6828 (2010).

22. D. Rueda, O. V. Boyarkin, and T. R. Rizzo, "Vibrational overtone spectroscopy of jet-cooled methanol from 5000 to 14 000 cm$^{-1}$," *J. Chem. Phys.* **122,** 044314 (2005).

23. J. P. Perchard, F. Romain, and Y. Bouteiller, "Determination of vibrational parameters of methanol from matrix-isolated infrared spectroscopy and ab initio calculations. Part 1 – Spectral analysis in the domain 11 000-200 cm$^{-1}$," *Chem. Phys.* **343,** 35-46 (2008).

24. S. W. Sharpe, T. J. Johnson, R. L. Sams, P. M. Chu, G. C. Rhoderick, and P. A. Johnson, "Gas-phase databases for quantitative infrared spectroscopy," *Appl. Spectrosc.* **58,** 1452-1461 (2004).

25. H. Yi, Q. Liu, A. J. Fleisher, and J. T. Hodges, "High-accuracy $^{12}C^{16}O_2$ line intensities in the 2 μm wavelength region measured by frequency-stabilized cavity ring-down spectroscopy," *J. Quant. Spectrosc. Radiat. Transfer* **206,** 367-377 (2017).

26. G.-W. Truong, K. O. Douglass, S. E. Maxwell, R. D. van Zee, D. F. Plusquellic, J. T. Hodges, and D. A. Long, "Frequency-agile, rapid scanning spectroscopy," *Nat. Photon.* **7,** 532-534 (2013).

27. J. T. Hodges, H. P. Layer, W. W. Miller, and G. E. Scace, "Frequency-stabilized single-mode cavity ring-down apparatus for high-resolution absorption spectroscopy," *Rev. Sci. Instrum.* **75,** 849 (2004).

28. G. C. Rhoderick, "Gas chromatographic analysis of methanol in gas mixtures," in *Measurement of Toxic and Related Air Pollutants*, (Air & Waste Management Association, United States, 1998) pp. 1034-1043.

29. T. F. Refaat, M. Petros, U. N. Singh, C. W. Antill, T.-H. Wong, R. G. Remus, K. Reithmaier, J. Lee, S. C. Bowen, B. D. Taylor, A. M. Welters, A. Noe, and S. Ismail, "Airborne direct-detection 2-μm triple-pulse IPDA lidar integration for simultaneous and independent atmospheric water vapor and carbon dioxide active remote sensing." Proc. SPIE *10779*, Lidar Remote Sensing for Environmental Monitoring XVI, 1077902 (2018).

30. J. Yu, M. Petros, U. Singh, T. Refaat, K. Reithmaier, R. Remus, and W. Johnson, "An airborne 2-μm double-pulsed direct-detection lidar instrument for atmospheric $CO_2$ column measurements," *J. of Atmos. and Ocea. Tech*. **34,** 385 (2017).

31. I. E. Gordon, L. S. Rothman, C. Hill, R. V. Kochanov, Y. Tan, P. F. Bernath, M. Birk, V. Boudon, A. Campargue, K. V. Chance, B. J. Drouin, J.-M. Flaud, R. R. Gamache, J. T. Hodges, D. Jacquemart, V. I. Perevalov, A. Perrin, K. P. Shine, M.-A. H. Smith, J. Tennyson, G. C. Toon, H. Tran, V. G. Tyuterev, A. Barbe, A. G. Császár, V. M. Devi, T. Furtenbacher, J. J. Harrison, J.-M. Hartmann, A. Jolly, T. J. Johnson, T. Karman, I. Kleiner, A. A. Kyuberis, J. Loos, O. M. Lyulin, S. T. Massie, S. N. Mikhailenko, N. Moazzen-Ahmadi, H. S. P. Müller, O. V. Naumenko, A. V. Nikitin, O. L. Polyansky, M. Rey, M. Rotger, S. W. Sharpe, K. Sung, E. Starikova, S. A. Tashkun, J. Vander Auwera, G. Wagner, J. Wilzewski, P. Wcisło, S. Yu, and E. J. Zak, "The HITRAN2016 molecular spectroscopic database," *J. Quant. Spectrosc. Radiat. Transfer* **203,** 3-69 (2017).

32. T. Shimanouchi, "Tables of molecular vibrational frequencies consolidated volume I," *Nat. Stand. Ref. Data Ser., Nat. Bur. Stand. (NSRDS-NBS)* **39,** 1-160, (1972).

33. J. J. Harrison and P. F. Bernath, "Infrared absorption cross sections for propane ($C_3H_8$) in the 3 μm region," *J. Quant. Spectrosc. Radiat. Transfer* **111,** 1282-1288 (2010).

34. R. V. Kochanov, I. E. Gordon, L. S. Rothman, K. P. Shine, S. W. Sharpe, T. J. Johnson, T. J. Wallington, J. J. Harrison, P. F. Bernath, M. Birk, G. Wagner, K. Le Bris, I. Bravo, and C. Hill, "Infrared absorption cross-sections in HITRAN2016 and beyond: Expansion for climate, environment, and atmospheric applications," *J. Quant. Spectrosc. Radiat. Transfer* **230,** 172-221 (2019).

35. D. Romanini, I. Ventrillard, G. Méjean, J. Morville, and E. Kerstel, "Introduction to Cavity Enhanced Absorption Spectroscopy," in *Cavity-Enhanced Spectroscopy and Sensing,* edited by G. Gagliardi and H.-P. Loock, (Springer, New York, 2014), pp 1-60.

36. P. R. Griffiths and J. A. de Haseth, *Fourier Transform Infrared Spectrometry*, 2$^{nd}$ edition, (John Wiley & Sons, Inc., Hoboken, New Jersey, 2007).

37. E. E. Whiting, "An empirical approximation to the Voigt profile," *J. Quant. Spectrosc. Radiat. Transfer* **8,** 1379-1384 (1968).

38. L. S. Rothman, D. Jacquemart, A. Barbe, D. C. Benner, M. Birk, L. R. Brown, M. R. Carleer, C. Chackerian Jr., K. Chance, L. H. Coudert, V. Dana, V. M. Devi, J.-M. Flaud, R. R. Gamache, A. Goldman, J.-M. Hartmann, K. W. Jucks, A. G. Maki, J.-Y. Mandin, S. T. Massie, J. Orphal, A. Perrin, C. P. Rinsland, M. A. H. Smith, J. Tennyson, R. N. Tolchenov, R. A. Toth, J. Vander Auwera, P. Varanasi, and G. Wagner, "The *HITRAN* 2004 molecular spectroscopic database," *J. Quant. Spectrosc. Radiat. Transfer* **96,** 139-204 (2005).